# Phase Distribution Properties of a Moving Five-Level (Inverted Y)-Type Atom in the Presence of a Non-linear Medium

*Sameh T. Korashy[1], Mohammed A. Saleem[2], T. M. El-Shahat[3]*

[1,2] *Department of Mathematics, Faculty of Science, Sohag University, 82524, Sohag, Egypt*

[3] *Department of Mathematics, Faculty of Science, Al-Azhar University, 71524, Assiut, Egypt*

**Abstract**  In this paper, we study the nonlinear kerr effect in an inverted Y-type five-level atom moving through an open cavity. Under realistic parameters, we show that the kerr-medium affects the atomic population inversion and the phase probability distribution. A remarkable feature of the present model is that the kerr nonlinearity with a multi-level atom allows for a large detection of the system dynamics. These results offer new possibilities for controlling the system dynamics using the time-dependent interaction.

## 1    Introduction

The Jaynes-Cummings model (JCM), a soluble fully quantum mechanical model of an atom in a field, was first use (in 1963) [1] to examine the interaction of a two-level atom with a single quantized mode of an optical cavity's electromagnetic field. It is widely use in cavity QED and circuit QED, while considering the rotating wave approximation [2]. Much attention has been paid to generalize the JCM in several fields such as multi-photon transition, multi-level atoms, intensity-dependent coupling, multi-atoms interaction, multi-mode fields, Stark shift and Kerr nonlinearity [3-15].

On the other hand, various efforts have been made to analyze multi-level atom in cavity field. One of the interesting examples is the system of three-level atom in different configurations interacting either with one- or two-mode field [3, 7, 16-21]. The entanglement as well as the geometric phase of the atom-field interaction have been carried out in [18-22], while the dynamics of a three-level interacting with a single-mode field in an optical cavity was performed in [23]. Another example of multi-level atoms interacting with the cavity field problems is the system of four-level [24-30]. For instance, the author in [26], has considered the quantum mutual entropy of a single four-level atom strongly coupled to a cavity field and driven by a laser field. Abdel-Aty et al. investigated an intensity coupling regime consisting of a $\lambda$-type four-level atom interacting with a single-mode quantized field [27]. More recently, we have proposed a non-linear time-dependent two two-level atoms system in [31, 32].

---

[1]    E-mail address: skorashe@yahoo.com
[2]    E-mail address: abuelhassan@yahoo.com
[3]    E-mail address: el_shahat@yahoo.com

A lot of researches focus on studying five-level atomic system in different areas of quantum optics. The optical switching by controlling the double dark resonance in an $N$-tripod five-level atom is explored in [33]. Colossal Kerr nonlinearity based on electromagnetically-induced transparency in a five-level double-ladder atomic system has been studied in [34]. The dynamical properties of a five-level fan type atom with a triple ground state interacting with one-mode electromagnetic cavity field in the presence of Kerr-like medium have been analyzed in [35]. The effect of a control field on the optical properties of a five-level inverted $Y$-type atomic system was discussed in [36]. The one- and two-dimensional atom localization behaviors via spontaneous emission in a coherently-driven five-level atomic system by means of a radiofrequency field driving a hyperfine translation, have been investigated in [37]. Theoretical study of electromagnetically induced transparency in a five-level atom and application to Doppler-broadened and Doppler-free Rb atoms has been conducted in [38]. In recent years, much attention has been focused on the properties of the four- and five-level atomic systems when time-dependent coupling with the field is considered [39-44]. More recently, the dynamics of a five-level (double $\Lambda$)-type atom interacting with two-mode field in a cross Kerr-like medium have been investigated in [45].

In this work, a five-level (inverted $Y$)-type atom interacting with a single-mode cavity field investigates the coupling parameter to be time-dependent. We report the atomic inversion as well as the properties of the phase. The rest of the paper is organized as follows: In Sec. 2, we present the model then we solve the system by considering approximations based on the rotating wave approximation. In Sec. 3, we report the atomic inversion and the dynamical properties for different regimes followed by a discussion of the reported numerical results. In Sec. 4, the phase properties are investigated. Finally, we provide concluding remarks in Sec. 5.

## 2  The Model

We consider a moving five-level (inverted $Y$)-type atom with the energy levels $\omega_5 > \omega_4 > \omega_3 > \omega_2 > \omega_1$ interacting with a single-mode quantized field of frequency $\Omega$ in an optical cavity surrounded by a Kerr nonlinear medium. The transitions $|5\rangle \leftrightarrow |4\rangle$, $|4\rangle \leftrightarrow |3\rangle$, $|3\rangle \leftrightarrow |1\rangle$, and $|3\rangle \leftrightarrow |2\rangle$ are allowed while the transition $|2\rangle \leftrightarrow |1\rangle$ is forbidden due to the selection rule (Figure. 1). Inspecting the system, we find that the five level atom consists of two sub-systems. In fact, states 1, 2 and 3 form a typical $\lambda$ configuration while omitting state 2 gives a four-state cascade-type system. In the Rotating-Wave Approximation (RWA), the system Hamiltonian is given by:

$$\hat{H} = \sum_{i=1}^{5} \omega_i \hat{\sigma}_{ii} + \Omega \hat{a}^\dagger \hat{a} + \gamma_1(t)(\hat{a}^k \hat{\sigma}_{13} + \hat{a}^{\dagger k} \hat{\sigma}_{31}) + \gamma_2(t)(\hat{a}^k \hat{\sigma}_{23} + \hat{a}^{\dagger k} \hat{\sigma}_{32})$$
$$+ \gamma_3(t)(\hat{a}^k \hat{\sigma}_{34} + \hat{a}^{\dagger k} \hat{\sigma}_{43}) + \gamma_4(t)(\hat{a}^k \hat{\sigma}_{45} + \hat{a}^{\dagger k} \hat{\sigma}_{54})$$
$$+ \chi \hat{a}^{\dagger 2} \hat{a}^2. \qquad (1)$$

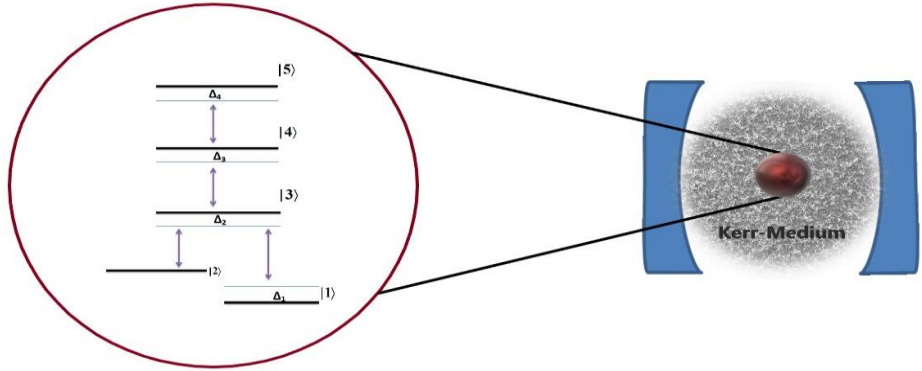

**Fig. 1** Schematic diagram of a five-level (inverted $Y$)-type atom interacting with a single-mode cavity field.

In seek of simplicity, we take $\hbar = 1$, $\hat{\sigma}_{ii} = |i\rangle\langle j|$ are the atomic raising or lowering operators. $\hat{a}^{\dagger}$ and $\hat{a}$, represent the field creation and annihilation bosonic operators, respectively. $\Omega$ designs the frequency of the mono-mode field. $\gamma_i(t)$, $i = 1,\ldots 4$ are the atom-field couplings. We denote by $\chi$ the third-order non-linearity of the Kerr-medium. The interaction Hamiltonian can be rewritten as:

$$\begin{aligned}\hat{H}_I &= \gamma_1(t)(\hat{a}^k e^{-i\Delta_1 t}\hat{\sigma}_{13} + \hat{a}^{\dagger k}e^{i\Delta_1 t}\hat{\sigma}_{31}) + \gamma_2(t)(\hat{a}^k e^{-i\Delta_2 t}\hat{\sigma}_{23} + \hat{a}^{\dagger k}e^{i\Delta_2 t}\hat{\sigma}_{32})\\ &+ \gamma_3(t)(\hat{a}^k e^{-i\Delta_3 t}\hat{\sigma}_{34} + \hat{a}^{\dagger k}e^{i\Delta_3 t}\hat{\sigma}_{43}) + \gamma_4(t)(\hat{a}^k e^{-i\Delta_3 t}\hat{\sigma}_{45} + \hat{a}^{\dagger k}e^{i\Delta_3 t}\hat{\sigma}_{54})\\ &+ \chi \hat{a}^{\dagger 2}\hat{a}^2,\end{aligned} \qquad (2)$$

In this study, the detuning are given by $\Delta_1 - \omega_3 - \omega_1 + k\Omega$ and $\Delta_j = \omega_{j+1} - \omega_j + k\Omega$ , for $j = 2,3,4$. Also, we consider $\gamma_1(t) = \gamma_2(t) = \gamma_3(t) = \gamma_4(t) = \gamma(t) = \lambda_j \cos(\mu t) = \frac{\lambda_j}{2}(e^{i\mu t} + e^{-i\mu t})$. Here, $\lambda$ is an arbitrary constant defining the time-dependent coupling torque of the matter–field interaction. In addition, we consider that the interaction intensity is not uniform denoted by $\mu$ the coupling variation parameter. As an approximation we neglect the rapidly oscillating (counter rotating) terms in the exponential. This approximation is very similar to the RWA and has been used extensively to solve such physical models. Hence, the interaction Hamiltonian becomes:

$$\begin{aligned}\hat{H}_I &= \frac{\lambda_1}{2}(\hat{a}^k e^{-i\delta_1 t}\hat{\sigma}_{13} + \hat{a}^{\dagger k}e^{i\delta_1 t}\hat{\sigma}_{31}) + \frac{\lambda_2}{2}(\hat{a}^k e^{-i\delta_2 t}\hat{\sigma}_{23} + \hat{a}^{\dagger k}e^{i\delta_2 t}\hat{\sigma}_{32})\\ &+ \frac{\lambda_3}{2}(\hat{a}^k e^{-i\delta_3 t}\hat{\sigma}_{34} + \hat{a}^{\dagger k}e^{i\delta_3 t}\hat{\sigma}_{43}) + \frac{\lambda_4}{2}(\hat{a}^k e^{-i\delta_4 t}\hat{\sigma}_{45} + \hat{a}^{\dagger k}e^{i\delta_4 t}\hat{\sigma}_{54})\\ &+ \chi \hat{a}^{\dagger 2}\hat{a}^2,\end{aligned} \qquad (3)$$

where $\delta_j = \Delta_j - \mu$ for $j = 1,..,4$. For $t > 0$ the wave function is given by:

$$\begin{aligned}|\Psi(t)\rangle &= \sum_{n=0}^{\infty} [A_1(n,t)|1,n\rangle + A_2(n,t)|2,n\rangle + A_3(n+k,t)|3,n+k\rangle\\ &+ A_4(n+2k,t)|4,n+2k\rangle + A_5(n+3k,t)|5,n+3k\rangle].\end{aligned} \qquad (4)$$

To derive the wave function, we consider the atom-field initial state as:

$$|\Psi(0)\rangle_{AF} = \sum_{n=0}^{\infty} q_{n+3k}|5,n+3k\rangle, \qquad (5)$$

where $q_n = e^{-|\alpha|^2/2} \frac{\alpha^n}{\sqrt{n!}}$, $|\alpha|^2 = \bar{n}$ is the initial mean photon number for the mode. Now, by substituting $|\Psi(t)\rangle$ from Eq. (4) and $\widehat{H}_I$ from Eq. (3) in the time-dependent Schrodinger equation $i\frac{\partial}{\partial t}|\Psi(t)\rangle = \widehat{H}_I |\Psi(t)\rangle$, we get the atomic probability amplitudes given by Eqs. (17) in the appendix.

Under the assumption of the conditions $\omega_1 = \omega_2$, $\Delta_1 = \Delta_2$ and $\lambda_1 = \lambda_2$ (that leads to $\alpha_1 = \alpha_2$), the probability amplitudes $A_1(n,t)$ and $A_2(n,t)$ satisfy similar differential equations leading to $A_1(n,t) = A_2(n,t)$. Consequently, the considered atom-field system has an appropriate analytical solution. Therefore, the system of the coupled differential equations for the probability amplitudes is written as:

$$i\frac{d}{dt}\begin{pmatrix} A_1 \\ A_3 \\ A_4 \\ A_5 \end{pmatrix} = \begin{pmatrix} \alpha_1 & v_1 e^{-i\delta_1 t} & 0 & 0 \\ 2v_1 e^{i\delta_1 t} & \alpha_3 & v_3 e^{-i\delta_3 t} & 0 \\ 0 & v_3 e^{i\delta_3 t} & \alpha_4 & v_4 e^{-i\delta_4 t} \\ 0 & 0 & v_4 e^{i\delta_4 t} & \alpha_5 \end{pmatrix} \begin{pmatrix} A_1 \\ A_3 \\ A_4 \\ A_5 \end{pmatrix}. \qquad (6)$$

The elements $A_i$ for i= 1,...,5 are solutions of Eqs. (6) (given in the Appendix). For $t > 0$, the density matrix describing the system is

$$\hat{\varrho}(t) = |\Psi(t)\rangle\langle\psi(t)| = \sum_{i,j=1}^{5} \varrho_{ij}(t),$$

where

$$\varrho_{11}(t) = \sum_{n,l=0}^{\infty} q_n q_l^* A_1(n,t) A_1^*(l,t) |n,1\rangle\langle l,1|$$
$$\varrho_{33}(t) = \sum_{n,l=0}^{\infty} q_n q_l^* A_3(n+k,t) A_3^*(l+k,t) |n+k,3\rangle\langle l+k,3|,$$
$$\varrho_{44}(t) = \sum_{n,l=0}^{\infty} q_n q_l^* A_4(n+2k,t) A_4^*(l+2k,t) |n+2k,4\rangle\langle l+2k,4|,$$
$$\varrho_{55}(t) = \sum_{n,l=0}^{\infty} q_n q_l^* A_5(n+3k,t) A_5^*(l+3k,t) |n+3k,5\rangle\langle l+3k,5|,\ldots,$$
$$\varrho_{ij}(t) = \varrho_{ji}^*(t). \qquad (7)$$

Now, we can easily find the reduced density operator for the field by tracing over the atomic variables. Then, the reduced density matrix of the cavity field takes the form: $\hat{\varrho}_F(t) = Tr_A[\hat{\varrho}(t)]$. We notice that, if $\mu = 0$ (time-independent case) the coupling parameters $\lambda$ should be replaced by $2\lambda$ in (17). In this case, we are dealing with the ordinary JCM for a five-level atomic system. In fact, we can not get the ordinary JCM case by simply letting $\mu = 0$, because we have to consider the slow and fast oscillatory terms into account. Note that, in the figures, the constants $\lambda_i = \lambda$ are real parameters and the interaction time is scaled to $\tau = \lambda t$.

## 3    Atomic Population Inversion

An important aspect of the atom-field interaction is the collapse and revival phenomena. Therefore, we investigate the dynamics of the atomic inversion which is defined as the difference between the population of the exited state $|5\rangle$ and the ground states. This is written as [3, 46]

$$W(\tau) = \sum_{n=0}^{\infty} [|A_5(n+3k,\tau)|^2 - 2|A_1(n,\tau)|^2]. \qquad (8)$$

In the following, the time-independent and the time-dependent cases are examined for one-photon ($k = 1$) and two-photon ($k = 2$) processes. Thus we report the influence of the time-dependent coupling, detuning and Kerr-medium on the atomic population inversion. The dynamics of the atomic population inversion is displayed in Figs. 2-4 for $\bar{n} = 20$. The left plots are reported for $k=1$ while the right plots are for $k=2$.

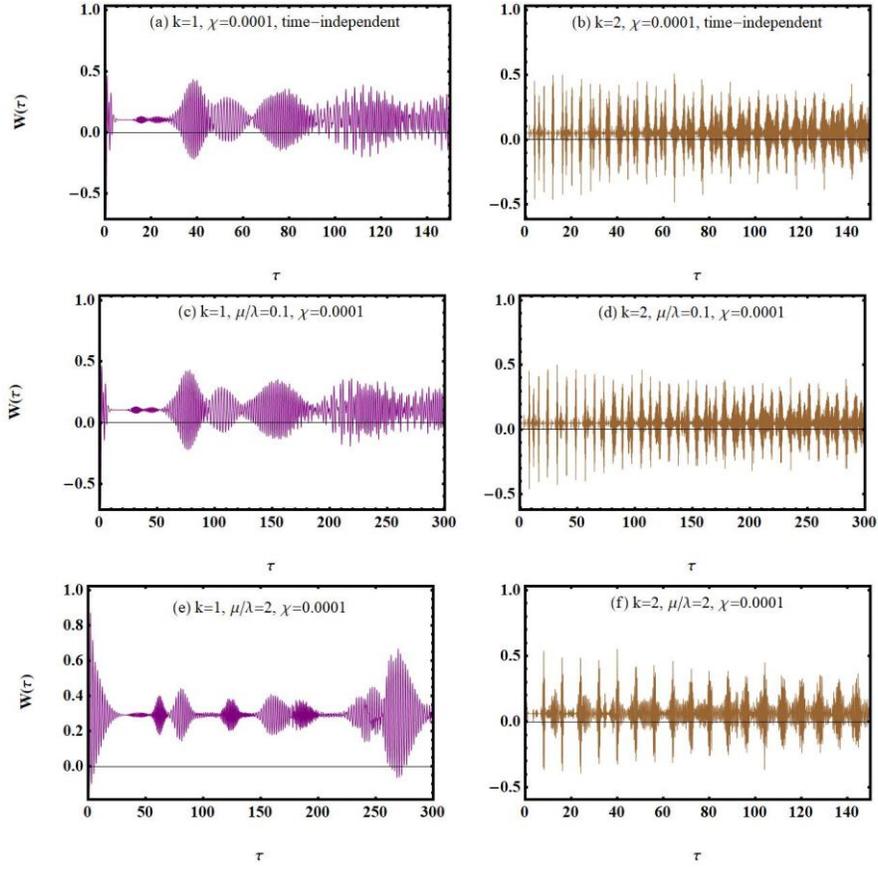

**Fig. 2** Evolution of the atomic population inversion $W(\tau)$ for a five-level atom interacting with a single-mode coherent field for $k=1$ (left plot), and for $k=2$ (right plot) for the parameters $\bar{n}=20$, $\Delta_1=\Delta_2=\Delta_4=0$, $\chi=0.0001$ and for: (a,b) time-independent, (c,d) $\mu/\lambda=0.1$, (e,f) $\mu/\lambda=2$.

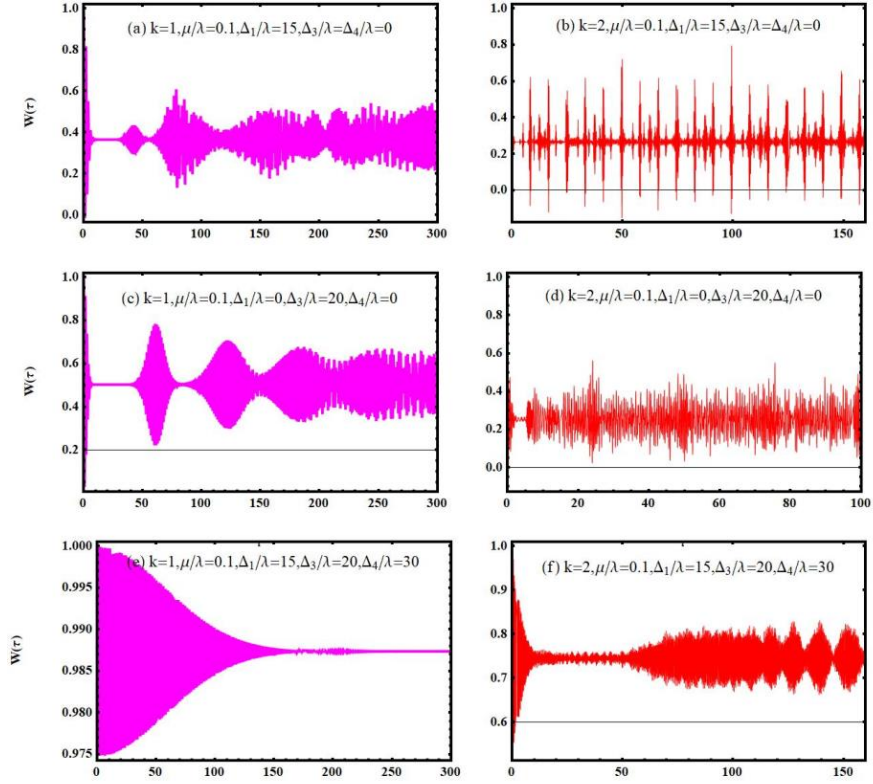

**Fig. 3** Evolution of the atomic population inversion $W(\tau)$ for a five-level atom interacting with a single-mode coherent field for $k = 1$ (left plot), and for $k = 2$ (right plot) for the parameters $\bar{n} = 20$, $\mu/\lambda = 0.1$, $\chi = 0.0001$ and for: (a,b) $\Delta_1/\lambda = 15, \Delta_2/\lambda = \Delta_4/\lambda = 0$, (c,d) $\Delta_1/\lambda = 0, \Delta_3/\lambda = 20, \Delta_4/\lambda = 0$, (e,f) $\Delta_1/\lambda = 15, \Delta_3/\lambda = 20, \Delta_4/\lambda = 30$.

In Fig. 2(a) and Fig. 2(b), we have considered the time-independent case in the absence of the detuning ($\Delta_1/\lambda = \Delta_3/\lambda = \Delta_4/\lambda = 0$ and $\chi = 0.0001$). The behavior of the atomic population inversion exhibits collapse and revival phenomena. The mean value of oscillations in Fig. 2(b) is less than that in Fig. 2(a). In Fig. 2(c) and Fig. 2(d), (where the time-dependent coupling $\mu/\lambda = 0.1$ is considered), $W(\tau)$ shows a similar behavior to the time-independent case ( except that, the time interval is elongated). By increasing the time-dependent coupling to $\mu/\lambda = 2$ in Fig. 2(e), the mean value of the collapses and the revivals is shifted upward. Also, we note that the behavior of the atomic population inversion has started with a set of revival with a big amplitude. The number of minima in Fig. 2(f) is less than the ones in Fig. 2(d). In Fig. 3, we examine the effect of the detuning on the atomic population inversion $W(\tau)$ for ($\mu/\lambda = 0.1$). Compared to Fig. 2(c) and Fig. 2(d), the mean value of oscillations is shifted upward clearly in Fig. 3(a) and Fig. 3(b). Indeed, In Fig. 3(b), the maximum value of oscillations is greater than that in Fig. 2(d). We also note that,in Fig. 3(c) , the mean value of oscillations is shifted upward and the first time interval of collapse increased compared to Fig. 3(a). The frequency of oscillations in Fig. 3(d) is also enhanced. This is also seen in Fig. 3(e) and Fig. 3(f) where the mean value of oscillations is higher than the previous cases. This means that the energy is enhanced in the atomic system. To discuss the influence of the Kerr-medium on the atomic population inversion in the presence of the time-dependent coupling($\mu/\lambda = 0.1$), and for n $\Delta_1/\lambda = \Delta_3/\lambda = \Delta_4/\lambda = 0$, we plot Fig. 4. For a small value of the Kerr-medium strength ($\chi = 0.01$), the behavior of $W(\tau)$ in Fig. 4(a) and Fig. 4(b) is similar to the behavior of $W(\tau)$ in Fig. 2(c) and Fig. 2(d). It is clear that, with the increase of the value of $\chi$, the behavior of $W(\tau)$ changes. The mean value of oscillations is shifted upward with some genres of periodicity in its behavior (see Fig. 4(c) and 4(d)). For bigger values of the Kerr-medium ($\chi = 1$), we have a higher positive mean value of oscillations and the maximum value of the fluctuations approaches one (see Fig. 4(e) and Fig. 4(f)).

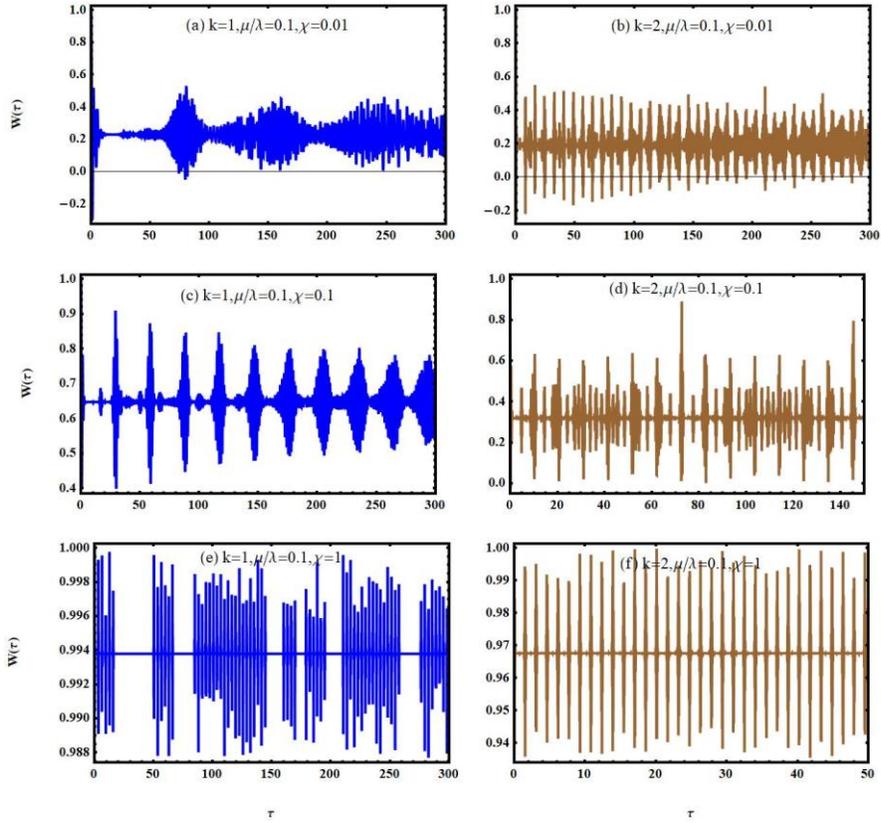

Fig. 4 Evolution of the atomic population inversion $W(\tau)$ for a five-level atom interacting with a single-mode coherent field for $k = 1$ (left plot), and for $k = 2$ (right plot) for the parameters $\bar{n} = 20$, $\mu/\lambda = 0.1$, $\Delta_1 = \Delta_3 = \Delta_4 = 0$, and for: (a,b) $\chi = 0.01$, (c,d) $\chi = 0.1$, (e,f) $\chi = 1$.

## 4  Phase Distribution Properties

In this section, we study and analyze the phase distribution (P-D) in the presence of the time-dependent coupling, detuning and Kerr-medium. The Hermitian phase operator has been defined by Pegg and Barnett [47-49]. Hence, the formulation of the Hermitian optical phase operator allows to examine rigorously the quantum phase properties of light. In their approach, they restricted the state space to $(s+1)$-dimensional space spanned by the first $(s+1)$ number states $|0\rangle, |1\rangle, \ldots, |s\rangle$ for a given mode of the field. All expectation values are calculated in the $(s+1)$-dimensional space. All orthonormal phase states are defined as follows

$$|\theta_m\rangle = \frac{1}{\sqrt{s+1}} \sum_{n=0}^{s} e^{in\theta_m} |n\rangle, \tag{9}$$

where $\theta_m = \theta_\circ + \frac{2m\pi}{s+1}$. $\theta_\circ$ is arbitrary and defines a particular basis set of $(s+1)$ mutually orthogonal phase states. Also, the Hermitian phase operator is defined as:

$$\hat{\phi}_\theta = \sum_{m=0}^{s} \theta_m |\theta_m\rangle\langle\theta_m|, \tag{10}$$

where the subscript $\theta$ indicates the dependence on the choice of $\theta_\circ$. The continuum phase distribution $P(\theta)$ is $P(\theta) = \lim_{s\to\infty} \frac{s+1}{2\pi} \langle\theta_m|\hat{\varrho}|\theta_m\rangle$ Replacing $\theta_m$ by the continuous phase variable $\theta$, we get then:

$$P(\theta, \tau) = \frac{1}{2\pi} \left[ 1 + 2Re \sum_{n,\iota:n>\iota}^{s} \varrho_{n,\iota}^F(\tau) e^{-i(n-\iota)\theta} \right], \tag{11}$$

where $\varrho_{n,\iota}^F(\tau)$ is the elements of the reduced density matrix of the cavity field. The expectation value of the phase operator moments in the state described by the density operator is given by

$$\langle \phi_\theta^k \rangle = \int_{\theta_\circ}^{\theta_\circ + 2\pi} \theta^k P(\theta) d\theta. \tag{12}$$

The phase variance which can be determined according to the formula below is of particular interest in the analysis of the phase properties of the field for the model under consideration

$$(\Delta\phi_\theta)^2 = \int_{-\pi}^{\pi} \theta^2 P(\theta) d\theta - (\langle\phi_\theta\rangle)^2. \tag{13}$$

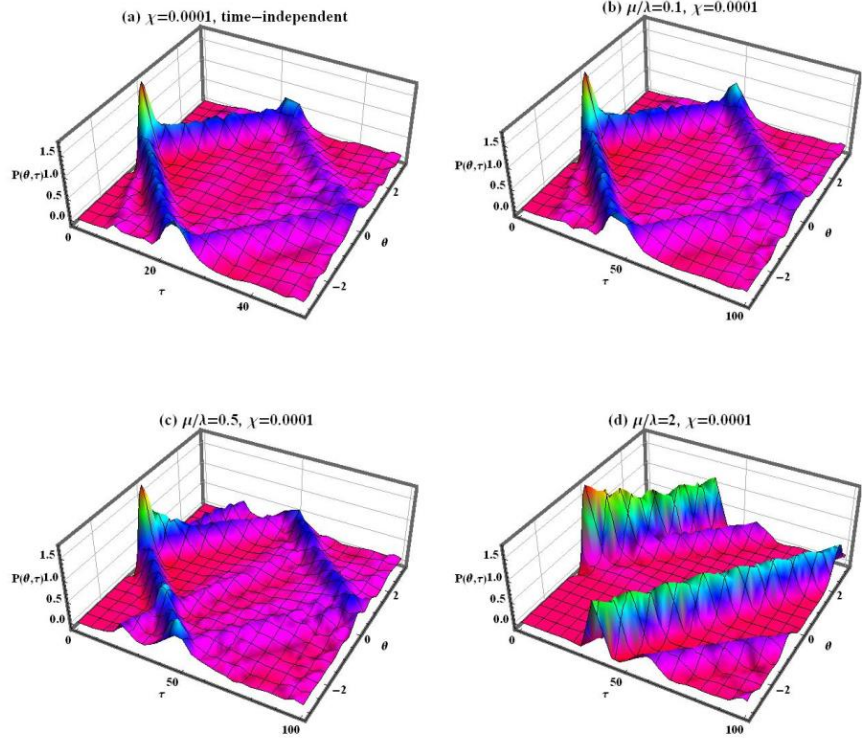

**Fig. 5** Evolution of the phase probability distribution $P(\theta,\tau)$ for a five-level atom interacting with a single-mode coherent field for $k=1$, and for the parameters $\bar{n}=5$, $\Delta_1=\Delta_3=\Delta_4=0$, $\chi=0.0001$ and for: (a) time-independent, (b) $\mu/\lambda=0.1$, (c) $\mu/\lambda=0.5$, (d) $\mu/\lambda=2$.

### 4.1 One-Photon Process

The phase probability distribution shows the field mode phase structure while the variance offers a clear understanding of the evolution of the fluctuations in the field phase. Hence, we will study both effects in this section. We split our work in two parts, first we start by analyzing the evolution of the phase probability distribution $P(\theta,\tau)$ versus the scaled time $\tau=\lambda t$ when $-\pi\leq\theta\leq\pi$. An illustration of the time evolution of the phase probability distribution for one-photon process ($k=1$) and for the initial mean photon numbers $\bar{n}=5$ is shown in Fig. 5 and Fig. 6. In Fig. 5(a), we consider the time-independent usual Jaynes-Cummings model for $\Delta_1=0.0001$ and for $\chi=0.0001$. we observe a single-peak structure corresponding to the initial coherent state. We recall that in the standard JCM [1], if the cavity field is initially in a coherent state and $\tau = 0$, then the phase distribution has a single peak structure corresponding to the initial coherent state. During the system's evolution, the single peak splits into two small peaks which gradually move away from each other. The peaks are symmetric about $\theta = 0$ so that the mean phase always remains equal to zero. The small peaks are further apart and reach the boundaries at $\theta = \pm\pi$. In addition, the peak maximum decreases and the small peaks converge until they intersect again at $\theta = 0$. Now, in Fig. 5(b), under the consideration of the time-dependent coupling $\mu/\lambda = 0.1$, we observe that there's no difference between the phase probability distribution behavior in the present case and that of the usual Jaynes-Cummings model except that there is an elongation in the time related to the collapse and revival phenomenon. When $\mu/\lambda = 0.5$, the asymmetric splitting starts to appear (see Fig. 5(c)). For $\mu/\lambda = 2$, the symmetric splitting disappears and new waves emerge(see Fig. 5(d)).

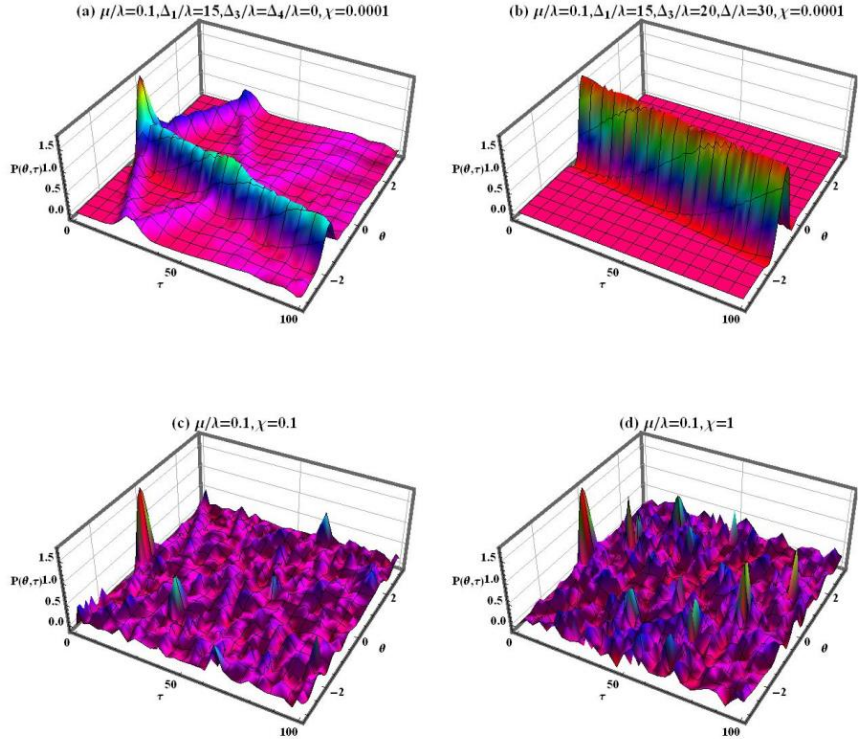

**Fig. 6** Evolution of the phase probability distribution $P(\theta, \tau)$ for a five-level atom interacting with a single-mode coherent field for $k = 1$, and for the parameters $\bar{n} = 5$, $\mu/\lambda = 0.1$, and for: (a) $\Delta_1/\lambda = 15$ $\Delta_3/\lambda = \Delta_4/\lambda = 0$, $\chi = 0.0001$, (b) $\Delta_1/\lambda = 15$, $\Delta_3/\lambda = 20$, $\Delta_4/\lambda = 30$, $\chi = 0.0001$, (c) $\chi = 0.1$, $\Delta_1 = \Delta_3 = \Delta_4 = 0$, (d) $\chi = 1$, $\Delta_1 = \Delta_3 = \Delta_4 = 0$.

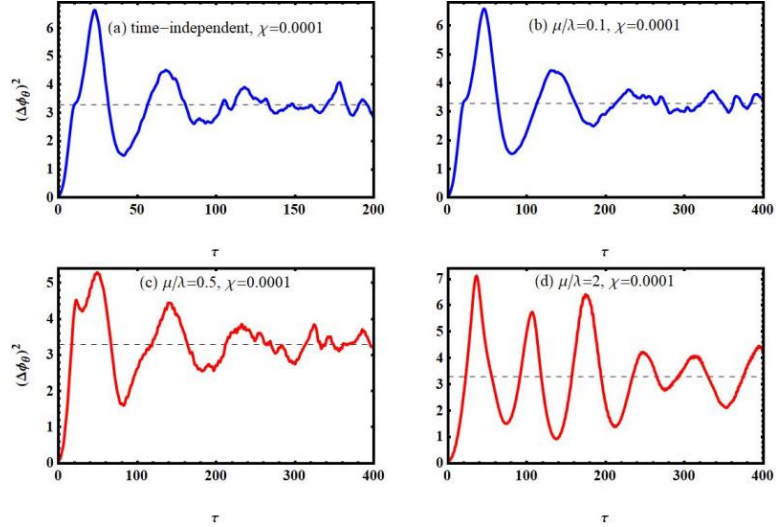

**Fig. 7** Dynamics of the variance of phase $(\Delta\phi_\theta)^2$ with the same conditions as stated in Fig. 5.

To explore the effect of the detuning parameters in the presence of the time-dependent coupling ($\mu/\lambda = 0.1$) and in the presence of Kerr-medium ($\chi = 0.0001$), we plot Fig. 6(a) and Fig. 6(b). In Fig 6(a), when $\Delta_1/\lambda = 15$, $\Delta_3/\lambda = \Delta_4/\lambda = 0$, the initial big peak is splitted into three types of peaks. The central peak is high. As the time progresses, other peaks are gone. When $\Delta_1/\lambda = 15$, $\Delta_3/\lambda = 20$, $\Delta_4/\lambda = 30$, the behavior of the phase probability distribution function is fully changed compared with the previous cases, we have only one strong peak starting from $\theta = 0$ and moving with its maximum value as the time evolves. When $\mu/\lambda = 0.1, \chi = 0.1$, and all the other parameters are zero, the situation is completely changed compared with Fig. 5(a) and Fig. 5(b). During the propagation of the coherent field in the Kerr-medium the phase distribution does not only shift, but also broadens. The larger the $\chi$, the stronger the phase diffusion which occurs except at some defined times (see Fig. 6(c) and Fig. 6(d)). In the second part of this analysis, we explore the evolution of the field phase fluctuations by investigating the time evolution of the second order variance $(\Delta\phi_\theta)^2$ against the scaled time $\tau$. For this purpose, we plot (in Fig. 7 and Fig. 8) for various values of $\mu/\lambda$, $\Delta_1/\lambda$, $\Delta_3/\lambda$, $\Delta_4/\lambda$, and $\chi$, and for one-photon process ($k = 1$). The considered mean photon number here is $\bar{n} = 5$). We notice that, the dynamic nature of $P(\theta, \tau)$ is observed through the behavior of $(\Delta\phi_\theta)^2$. Fig. 7(a), shows the time-independent scenario, we find that $(\Delta\phi)^2$ oscillates about $\frac{\pi^2}{3}$. This is identical to the JCM behavior. By considering the time-dependent coupling $\mu/\lambda = 0.1$, $(\Delta\phi_\theta)^2$ demonstrates the same dynamics as for the time-independent case except that the time interval is elongated (see Fig. 7(b)). We observe that the maximum value of oscillations is reduced for $\mu/\lambda = 0.5$, but when the value of $\mu/\lambda$ approaches 2, the maximum value of oscillations is increased again (see Fig. 7(c) and Fig. 7(d)). The dynamic of $(\Delta\phi_\theta)^2$ in the presence of detuning parameters, is reported in Fig. 8(a) and Fig. 8(b) for $\mu/\lambda = 0.1$. In Fig.8(b) we see a single maximum value in the considered time interval. (That reflects the behavior of $P(\theta, \tau)$ shown in Fig. 6(b)). The effect of the Kerr-medium on $(\Delta\phi_\theta)^2$ in the presence of the time-dependent coupling parameter ($\mu/\lambda = 0.1$) and in the absence of detuning parameters is plotted in Fig. 8(c) and Fig. 8(d). The dynamic of $(\Delta\phi_\theta)^2$ is completely changed in comparison with the previous cases, but still oscillates around $\frac{\pi^2}{3}$. For a large value of Kerr-medium, there is a remarkable change in the behavior of $(\Delta\phi_\theta)^2$. The frequency of oscillations is enhanced and a periodic behavior emerges (see Fig. 8(d)).

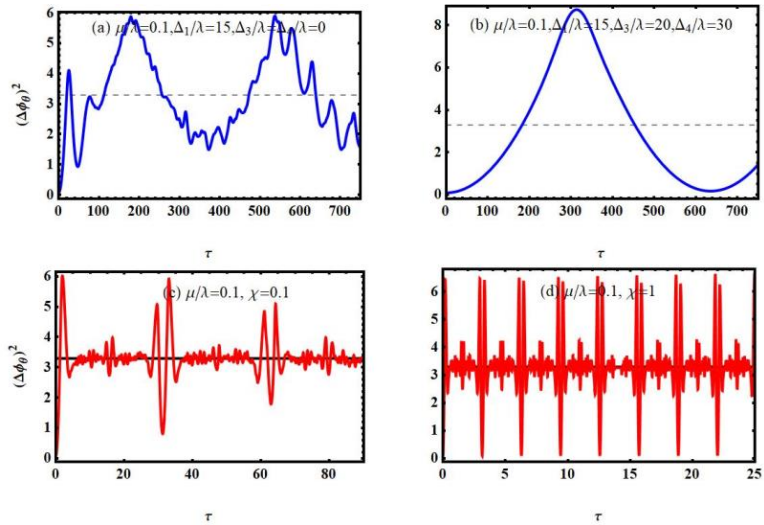

**Fig. 8** Dynamics of the variance of phase $(\Delta\phi_\theta)^2$ with the same conditions as stated in Fig. 6.

## 4.2 Two-Photon Process

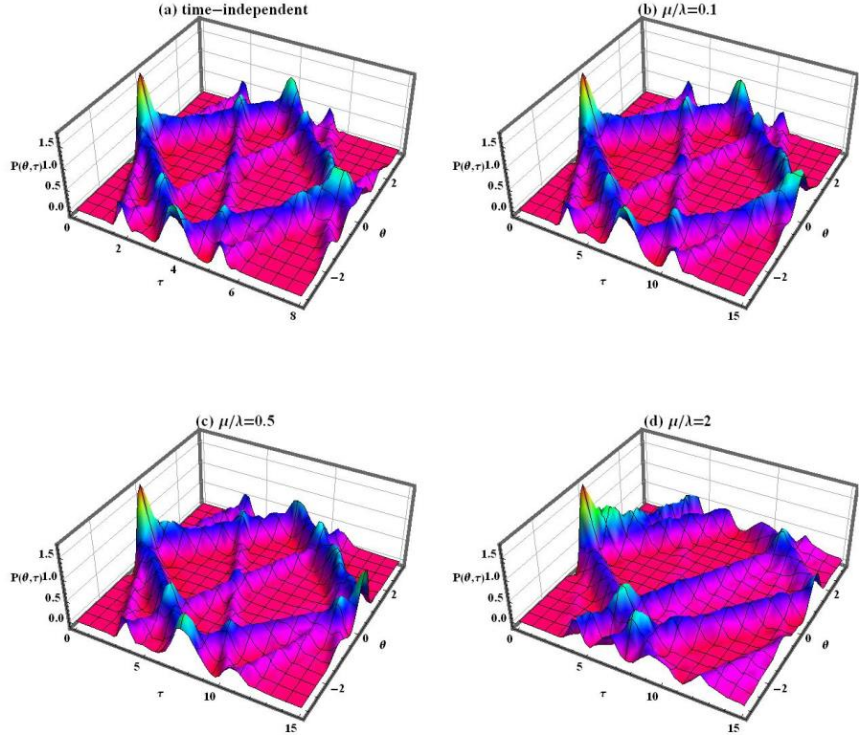

**Fig. 9** Dynamics of the phase probability distribution $P(\theta,\tau)$ with the same conditions as stated in Fig. 5, but for $k=2$.

In Fig. 9 and Fig. 10 we plot $P(\theta,\tau)$ versus the scaled time $\tau$ and $-\pi \leq \theta \leq \pi$ for the initial mean photon numbers $\bar{n}=5$, two-photon process ($k=2$) and for different values of the parameters $\mu/\lambda$, $\Delta_1/\lambda, \Delta_3/\lambda$, $\Delta_4/\lambda$, $\chi$. In fig.9 (a), we consider the time independent case and $\Delta_1 = \Delta_2 = \Delta_3 = \chi = 0.0001$, we observe a single-peak structure corresponding to the initial coherent state similar to that in Fig. 5(a). But, during the evolution of the system the single-peak splits up into four peaks. The first two peaks are lower than the others and move away from each other gradually. Moreover, the first peaks are symmetric around $\theta = 0$ and meet the boundaries at $\theta = \pm\pi$ before the others. Small peaks appear at $(\theta = 0)$. At Fig. 9(b), for the time-dependent coupling $\mu/\lambda = 0.1$, there is no essential difference between the phase probability distribution behavior for the usual Jaynes-Cummings model (time-independent case) and the present case, except that we observe an elongation in the considered time interval.

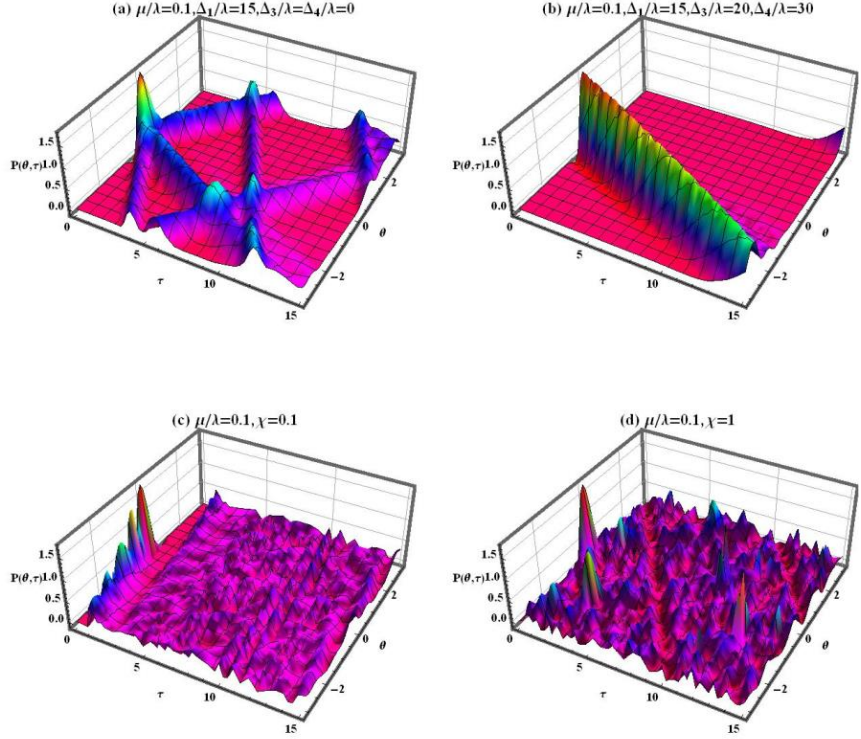

**Fig. 10** Dynamics of the phase probability distribution $P(\theta,\tau)$ with the same conditions as stated in Fig. 6, but for $k=2$.

The symmetric splitting is still present for $\mu/\lambda = 0.5$, but after $\theta = 0$ the four peaks start to merge. When $\mu/\lambda = 2$ the phase probability behavior is totally modified relative to the previous cases, the symmetric splitting disappears as well as one of the first two peaks. Also, peaks at the end of the considered time interval did meet far from the value $\theta = 0$. To investigate the influence of the detuning and Kerr-medium on the behavior of the phase probability distribution function $P(\theta,\tau)$ in the presence of the time-dependent coupling parameter, we plot Fig. 10. In Fig. 10(a), $\mu/\lambda = 0.1$, $\Delta_1/\lambda = 15$, $\Delta_3/\lambda = \Delta_4/\lambda$, and $\chi = 0.0001$. We notice that the dynamic of $P(\theta,\tau)$ is changed. There is an asymmetric splitting and one of the first two peaks that appears in Fig. 9(b) is disappeared. In addition, for $\mu/\lambda = 0.1$, $\Delta_1/\lambda = 15$, $\Delta_3/\lambda = 20$, $\Delta_4/\lambda = 30$, and $\chi = 0.0001$, the behavior of $P(\theta,\tau)$ is completely different from its dynamics in Fig. 9(b) and Fig. 10(a). We have only one peak starting from $\theta = 0$ which is reduced as the time evolves. At the end of the considered time interval, the peak is far from the value $\theta = 0$ (see Fig. 10(b)). When $\mu/\lambda = 0.1$, $\chi = 0.1$,

$\Delta_1 = \Delta_3 = \Delta_4 = 0$, we have only one peak which decreases as the time progresses. The remaining of the curve surface behaves like a small wave (see Fig. 10(c)). It is worth noting that, for $\mu/\lambda$ =0.1, $\chi$ =1, new small peaks appears (see Fig. 10(d)).

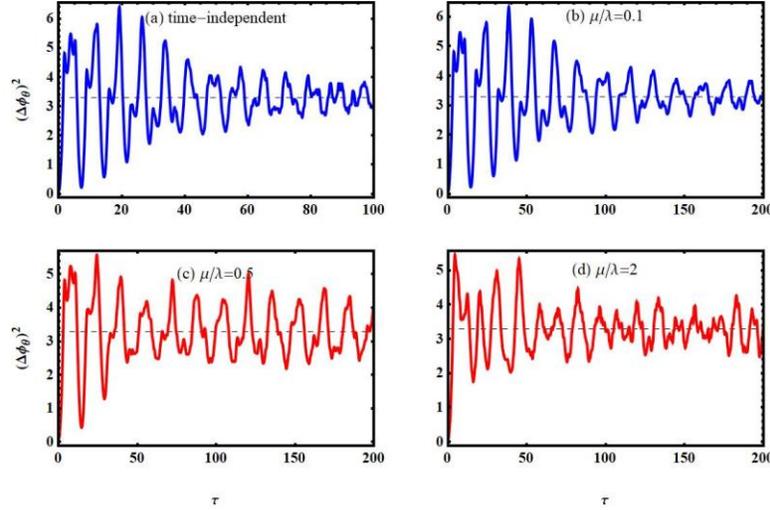

**Fig. 11** Dynamics of the variance of phase $(\Delta\phi_\theta)^2$ with the same conditions as stated in Fig. 7, but for $k = 2$.

Fig. 11 and Fig. 12 show the plots of the phase variance $(\Delta\phi_\theta)^2$ against the scaled time $\tau$ for two-photon process ($k = 2$) and for the same values of parameters as those taking in Fig. 9 and Fig. 10. In Fig. 11(a), we consider the time-independent case. For $k = 2$, $\Delta\phi_\theta^2$ oscillates around $\frac{\pi^2}{3}$. Which is similar to the case with the usual JCM model. In Fig 11, the frequency of oscillations is enhanced compared to the ones in Fig. 7. Also, the maximum value of $(\Delta\phi_\theta)^2$ in Fig. 11(d) is less than that in Fig. 7(d), when $\mu/\lambda = 2$. The behavior of $(\Delta\phi_\theta)^2$ in the presence of detunigs for $k = 2$ is displayed in Fig. 12(a) and Fig. 12(b) in the presence of the time-dependent coupling ($\mu/\lambda = 0.1$). The behavior of $(\Delta\phi_\theta)^2$ in Fig. 12(b) shows a single peak in a short time interval compared with that in Fig. 8(b). The maximum of this peak decreases dramatically as the time evolves. This is similar to the behavior of $P(\theta,\tau)$. The effect of the Kerr-medium on $(\Delta\phi_\theta)^2$ in the presence of the time-dependent coupling ($\mu/\lambda = 0.1$) and in the absence of detunings is plotted in Fig. 12(c) and Fig. 12(d).

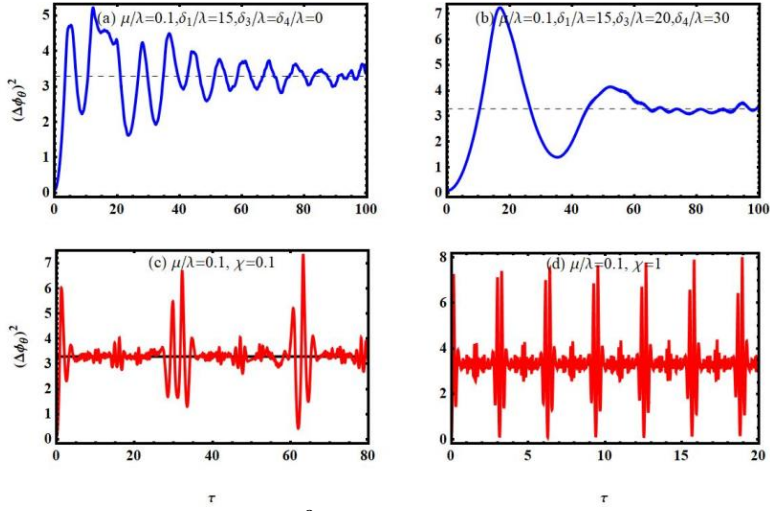

**Fig. 12** Dynamics of the variance of phase $(\Delta\phi_\theta)^2$ with the same conditions as stated in Fig. 8, but for $k=2$.

# 5  Conclusion

In this paper, we have invoked a new approach for the analysis of a general multi-level atomic system taking into account both the time-dependent interaction and the nonlinear medium. We have therefore investigated the effect of the nonlinear medium with different numbers of photons where we demonstrate that the non-linearity due to the kerr effect dramatically improves population inversion and contributes to an increase in phase space diffusion. Furthermore, a periodic behavior emerges in the dynamics of the phase space fluctuation due to the kerr effect. The time-dependent coupling leads to a time elongation in the dynamics of the atomic population inversion as well as in the phase space fluctuations. From another side, it controls the symmetric or asymmetric splitting in the phase probability distribution. Our treatment can handle both short as well as long periods of evolution, making it an appropriate tool for analyzing long-lived dynamics with future interest to analyze large numbers of atomic systems.

# 6  Appendix

By substituting $|\Psi(t)\rangle$ from Eq. (5) and $\widehat{H}_I$ from Eq. (4) in the time-dependent Schrodinger equation $i\frac{\partial}{\partial t}|\Psi(t)\rangle = \widehat{H}_I\,|\Psi(t)\rangle$, we get the following system of the differential equations for the atomic probability amplitudes:

$$i\frac{d}{dt}\begin{pmatrix}A_1\\A_2\\A_3\\A_4\\A_5\end{pmatrix}=\begin{pmatrix}\alpha_1 & 0 & v_1 e^{-i\delta_1 t} & 0 & 0\\ 0 & \alpha_2 & v_2 e^{-i\delta_2 t} & 0 & 0\\ v_1 e^{i\delta_1 t} & v_2 e^{i\delta_2 t} & \alpha_3 & v_3 e^{-i\delta_3 t} & 0\\ 0 & 0 & v_3 e^{i\delta_3 t} & \alpha_4 & v_4 e^{-i\delta_4 t}\\ 0 & 0 & 0 & v_4 e^{i\delta_4 t} & \alpha_5\end{pmatrix}\begin{pmatrix}A_1\\A_2\\A_3\\A_4\\A_5\end{pmatrix} \qquad (14)$$

where all the variables are given by:
$$\alpha_1 = \alpha_2 = \chi n(n-1), \quad \alpha_3 = \chi(n+k)(n+k-1),$$
$$\alpha_4 = \chi(n+2k)(n+2k-1), \quad \alpha_5 = \chi(n+3k)(n+3k-1),$$
$$v_1 = \frac{\lambda_1}{2}\sqrt{\frac{(n+k)!}{n!}}, \quad v_2 = \frac{\lambda_2}{2}\sqrt{\frac{(n+k)!}{n!}},$$
$$v_3 = \frac{\lambda_3}{2}\sqrt{\frac{(n+2k)!}{(n+k)!}}, \quad v_4 = \frac{\lambda_4}{2}\sqrt{\frac{(n+3k)!}{(n+2k)!}}. \tag{15}$$

Under the assumption of $\lambda_1 = \lambda_2$ the probability amplitudes $A_1(n,t)$ and $A_2(n,t)$ satisfy similar differential equations leading to $A_1(n,t) = A_2(n,t)$. Therefore, the probability amplitudes are given by:

$$i\frac{d}{dt}\begin{pmatrix}A_1\\A_3\\A_4\\A_5\end{pmatrix} = \begin{pmatrix}\alpha_1 & v_1 e^{-i\delta_1 t} & 0 & 0\\ 2v_1 e^{i\delta_1 t} & \alpha_3 & v_3 e^{-i\delta_3 t} & 0\\ 0 & v_3 e^{i\delta_3 t} & \alpha_5 & v_4 e^{-i\delta_4 t}\\ 0 & 0 & v_4 e^{i\delta_4 t} & \alpha_5\end{pmatrix}\begin{pmatrix}A_1\\A_3\\A_4\\A_5\end{pmatrix}. \tag{16}$$

Solving the above equation leads to:
$$A_1(n,t) = A_2(n,t) = \sum_{j=1}^{4} B_j e^{i\zeta_j t},$$
$$A_3(n+k,t) = -\frac{1}{v_1}\sum_{j=1}^{4} B_j(\alpha_1 + \zeta_j) e^{i(\zeta_j + \delta_1)t},$$
$$A_4(n+2k,t) = \frac{1}{v_1 v_3}\sum_{j=1}^{4} B_j[(\Gamma_1 + \zeta_j)(\alpha_1 + \zeta_j) - 2v_1^2] e^{i(\zeta_j + \delta_1 + \delta_3)t}, \tag{17}$$
$$A_5(n+3k,t) = \frac{1}{v_1 v_3 v_4}\sum_{j=1}^{4} B_j[(\alpha_1 + \zeta_j)[v_3^2 - (\Gamma_1 + \zeta_j)(\Gamma_2 + \zeta_j)] + 2v_1^2(\Gamma_2 + \zeta_j)]e^{i(\zeta_j + \delta)t}.$$

Applying the initial conditions for the atom and the field and using the above derived solutions, the $B_j$ coefficients read:
$$B_j = \frac{-q_{n+3k} v_1 v_3 v_4}{\zeta_{jk}\zeta_{jl}\zeta_{jm}}, \quad j \neq k \neq l \neq m = 1, 2, 3, 4, \tag{18}$$
where $\zeta_{jk} = \zeta_j - \zeta_k$, $\zeta_j$, (j=1,2,3,4) are the roots of the equation
$$\zeta^4 + a_1\zeta^3 + a_2\zeta^2 + a_3\zeta + a_4 = 0, \tag{19}$$

$$a_4 = \Gamma_3\Gamma_8 - \alpha_1\Gamma_1 v_4^2 + 2v_1^2 v_4^2,$$
$$a_3 = \Gamma_3\Gamma_7 + \Gamma_8 - v_4^2\Gamma_4,$$
$$a_2 = \Gamma_3\Gamma_6 + \Gamma_7 - v_4^2,$$
$$a_1 = \Gamma_3 + \Gamma_6, \quad \delta = \delta_1 + \delta_3 + \delta_4. \tag{20}$$

And,
$$\Gamma_1 = \alpha_3 + \delta_1, \quad \Gamma_2 = \alpha_4 + \delta_1 + \delta_3,$$
$$\Gamma_3 = \alpha_5 + \delta, \quad \Gamma_4 = \alpha_1 + \Gamma_1,$$
$$\Gamma_5 = \alpha_1\Gamma_1 - 2v_1^2, \quad \Gamma_6 = \Gamma_2 + \Gamma_4,$$
$$\Gamma_7 = \Gamma_5 + \Gamma_2\Gamma_4 - v_3^2, \quad \Gamma_8 = \Gamma_2\Gamma_5 - \alpha_1 v_3^2. \tag{21}$$

The four roots of the quartic equation (19) have [50] the following form
$$\zeta_{1(2)} = \frac{-a_1}{4} - \frac{1}{2}\sqrt{y_2 + \frac{y_1}{3d_2} + \frac{d_2}{3}} \mp \sqrt{z_2 - \frac{z_3}{4z_1}},$$

$$\zeta_{3(4)} = \frac{-a_1}{4} + \frac{1}{2}\sqrt{y_2 + \frac{y_1}{3d_2} + \frac{d_2}{3}} \mp \sqrt{z_2 + \frac{z_3}{4z_1}}, \tag{22}$$

where

$$z_1 = \sqrt{y_2 + \frac{y_1}{3d_2} + \frac{d_2}{3}}, \ z_2 = 2y_2 - \frac{y_1}{3d_2} - \frac{d_2}{3}, \ z_3 = -8a_3 + 4a_1a_2 - a_1^3,$$

$$y_1 = 12a_4 + a_2^2 - 3a_1a_3, \ y_2 = \frac{-2a_2}{3} + \frac{a_1^2}{4},$$

$$d_1 = 27a_3^2 - 72a_2a_4 + 2a_2^3 - 9a_1a_2a_3 + 27a_1^2a_4,$$

$$d_2 = \left(\frac{d_1 + \sqrt{d_1^2 - 4y_1^3}}{2}\right)^{\frac{1}{3}}. \tag{23}$$